# Pliant: Leveraging Approximation to Improve Datacenter Resource Efficiency

Neeraj Kulkarni, Feng Qi, and Christina Delimitrou
Cornell University
{nsk49,fq26,delimitrou}@cornell.edu

## Abstract

Cloud multi-tenancy is typically constrained to a single interactive service colocated with one or more batch, low-priority services, whose performance can be sacrificed when deemed necessary. Approximate computing applications offer the opportunity to enable tighter colocation among multiple applications whose performance is important. We present Pliant, a lightweight cloud runtime that leverages the ability of approximate computing applications to tolerate some loss in their output quality to boost the utilization of shared servers. During periods of high resource contention, Pliant employs incremental and interference-aware approximation to reduce contention in shared resources, and prevent QoS violations for co-scheduled interactive, latency-critical services. We evaluate Pliant across different interactive and approximate computing applications, and show that it preserves QoS for all co-scheduled workloads, while incurring a 2.1% loss in output quality, on average.

## 1. Introduction

Cloud computing has reached proliferation by offering *resource flexibility* and *cost efficiency* [11, 7, 23, 10]. Resource flexibility is achieved by users elastically scaling their resources on-demand, and releasing them when they no longer need them. Cost efficiency is achieved through multi-tenancy, i.e., by co-scheduling multiple jobs on the same physical platform to increase server utilization. Unfortunately multi-tenancy also leads to unpredictable performance, due to interference in shared resources [16, 36, 42, 17, 33, 20, 19, 32, 43, 44, 60, 50, 54, 18, 27, 45]. When the applications that suffer from interference are high priority, interactive services, such as websearch and social networking, multi-tenancy is disallowed altogether, hurting utilization, or - at best - interactive services are co-scheduled with low priority, best-effort workloads [33, 45, 51]. The performance of these workloads can be sacrificed at runtime to avoid performance degradation for the high priority service [51, 28, 17, 33, 42, 41, 26, 25]. Unfortunately this limits the options cloud providers have in terms of applications they can co-schedule with an interactive, latency-critical service. Approximate computing offers the potential to break this utilization versus performance trade-off in shared clouds.

Approximate computing applications include workloads from several fields, such as computer vision, machine learning, analytics, and scientific applications, and have the common feature that they can tolerate some loss in output accuracy in return for improved performance and/or energy efficiency [49, 21, 39, 38, 13]. Several cloud workloads fall under this category, such as big data analytics and ML applications, where achieving the highest output quality is often less important than latency and/or throughput. Exposing the knob of approximation to the cloud scheduler allows the system to sacrifice some accuracy in applications that can tolerate it, to preserve the services' quality-of-service (QoS) constraints.

We present Pliant, an online cloud runtime system that achieves both high QoS and high utilization by leveraging the ability of approximate computing applications to tolerate some loss in their output quality. Pliant enables aggressive co-scheduling of interactive, latency-critical services with approximate applications. Unlike prior cluster schedulers, Pliant does not consider applications co-scheduled with interactive services as low-priority, and does not penalize their performance [51, 18, 33]. Instead, a user expresses an application's tolerable approximation threshold to the scheduler, and Pliant dynamically adjusts approximation to the minimum amount needed to satisfy the tail latency QoS of the interactive service at each point in time, without exceeding this threshold.

Pliant consists of a lightweight performance monitor and an online dynamic compilation system. The monitor uses adaptive sampling of end-to-end latency to continuously check for QoS violations in the interactive service, while the compilation system is based on DynamoRIO [2], and adjusts the approximation degree of an application online. When the interactive service violates its QoS, Pliant first employs approximation to reduce interference in shared resources, and when that is not sufficient, it additionally reclaims one or more cores from the approximate application(s), yields them to the interactive service, and adjusts the approximation degree to ensure that the execution time of the approximate application(s) does not degrade. Pliant reclaims cores incrementally to guarantee that the approximate application only sacrifices the minimum amount of accuracy needed at each point in time, and leverages lightweight Linux signals to switch between approximation degrees, to avoid high overheads from dynamic compilation.

We evaluate Pliant on servers with 44 physical cores, with three popular open-source interactive services; memcached, a distributed in-memory cache [1], NGINX, a front-end web server [5], and MongoDB, a stateful persistent database [3]. We additionally use a set of 24 scientific applications from PARSEC [12], SPLASH2 [53], BioPerf [24], and Minebench [40] as the approximate applications. We show that Pliant is able to preserve both the tail latency QoS of the interactive services, and the nominal execution time of the approximate applications, with a 2.1% loss of output quality on average, and 5% loss in the worst case. In comparison, running the applications in precise mode results in a $2-10x$ increase in the tail latency of memcached, a dramatic degradation for latency-sensitive, interactive services. Finally, we explore the sensitivity of Pliant to the decision granularity, across input loads.

## 2. Related Work

We now review relevant work in interference-aware scheduling, approximate computing, and dynamic instrumentation.

**Contention-aware scheduling:** Sharing resources to increase utilization results in performance degradation [36, 16, 17, 43], and in some cases security vulnerabilities [19, 46, 57, 59, 58]. Several systems that aim to minimize destructive interference disallow colocation of jobs that contend in the same resources [16, 17, 37, 36, 42, 43], or partition resources to improve isolation [33, 19, 26, 25]. For example, BubbleFlux determines how the memory sensitivity of applications evolves over time, and prevents multiple memory-intensive services from sharing the same platform [36]. Similarly, DeepDive identifies the interference colocated VMs experience, and manages it transparently to the user [43]. Paragon [16] and Quasar [17] are cluster managers that leverage a set of practical online data mining techniques to determine the resource requirements of incoming cloud applications, and schedule them in a way that minimizes resource contention. In the same spirit, Nathuji et al. [42] develop Q-Clouds, a QoS-aware control framework that dynamically adjusts resource allocations to mitigate interference in virtualized clouds.

On the isolation front, Lo et al. [33] study the sensitivity of Google applications to different sources of interference, and combine hardware and software isolation techniques to preserve the QoS of latency-critical applications running alongside batch, low-priority workloads. Similarly Kasture et al. [26] implement fine-grain cache partitioning, and power allocation with RAPL [25] on servers that host one interactive, and one or more best-effort services. In all cases, a server hosts at most one high priority application; any remaining workloads are best effort, and their performance can be sacrificed when needed.

**Approximate computing techniques:** Finding the approximation potential of popular application classes, and generating language constructs to express and verify approximation has generated a large amount of related work. Carbin et al. [13] present language constructs for specifying acceptability properties in approximate programs, and develop a system that enables developers to obtain fully machine-checked verifications of their approximate applications. Sampson et al. [48] propose annotating data types that can be approximated, and automatically mapping such variables to low-power storage, and using low-power operations on them. They extend this work to map such variables to approximate storage devices in [49]. The same authors develop ACCEPT [47], a programmer-guided compiler framework that identifies approximable code, and automatically chooses the best approximation strategies for it. Finally, Misailovic et.al [39] present Chisel, an optimization framework that automatically generates approximate instructions and data that can be stored in approximate memory to improve energy efficiency, at the cost of some reliability and accuracy loss. They also propose compiler-level, accuracy-aware transformations that automatically generate approximate versions of programs [38]. A lot of this prior work focuses on improving the programmability and ease of development of approximate applications, to avoid tasking the user with generating approximate variants manually [39, 38, 48].

**Dynamic recompilation:** Open-source tools like DynamoRIO [2] enable online code transformations that can be used, among other reasons, to reduce the amount of resource contention the instrumented application incurs in a multi-tenant system. For example, Protean Code [28] is a co-designed compiler and runtime built on top of LLVM that enables compiler transformations at runtime with less than 1% overhead. The runtime dynamically mitigates cache pressure via fine-grain code transformations that disable prefetching for the low-priority application during periods of high resource contention. There are also several dynamic optimization systems that do not directly aim to reduce resource contention, but focus on code transformations that optimize application performance. Mojo [15] was the first tool to facilitate dynamic software optimizations on an x86 architecture, while ADAPT [52] is a compiler-supported high-level adaptive optimization system, which leverages user-provided optimizations and heuristics to efficiently explore the application design space at runtime. Finally, ADORE [34, 35] is a dynamic optimization runtime that monitors performance counters during application execution to detect hotspots, and leverages online compilation, similar to DynamoRIO, to tune data cache prefetching. Trident [55] builds on ADORE, and uses hardware support to reduce the overheard of online profiling.

## 3. Approximation Design Space Exploration

We first examine the potential approximation offers in trading off output quality for performance QoS and efficiency in multi-tenant cloud settings. We explore several approximation strategies whose performance and efficiency benefits have been previously shown [48, 39], including *loop perforation*, *synchronization elision*, and *lower precision data types*.

- **Loop perforation:** This technique omits a fraction of the iterations of a loop. Typical approximate computing applications, like analytics and machine learning workloads are iterative in nature, making loop perforation a good candidate for approximation. There are multiple ways to perforate a loop. For example, to reduce a loop by a factor $p$, we can execute only a chunk of (MAX_ITER/p) iterations, or execute every $p^{th}$ iteration. We can also reduce the loop by a factor of $(p-1)/p$ by not executing every $p^{th}$ iteration. Perforating a loop lowers the accuracy of the output, however, it also leads to lower execution time, and reduced memory traffic by avoiding the data accesses of the omitted iterations. Note that the decrease in output quality is not always proportional to the decrease in execution time. This is because, depending on each application's logic, different loop iterations may contribute differently to output quality. For example, in the simulated annealing algorithm used in `canneal` [12], if the cost of the randomly chosen neighboring solution is not greater than the current solution, the current solution is retained and no useful work is done. Omitting such iterations decreases execution time without any observable impact on quality.
- **Synchronization elision:** Synchronization constructs, like locks and barriers, which are used to guarantee correctness can be elided at the cost of some inaccuracy in the final result. Removing locks reduces the memory traffic that acquiring locks incurs, which can be significant, especially for highly contended locks. Apart from memory traffic, synchronization elision also benefits performance, as threads do not wait to synchronize, shortening execution time.
- **Lower precision:** This technique leverages the ability of certain applications to operate with lower precision to replace high-precision variables such as "double" with lower precision data types, like "float" and "int". Reducing precision reduces both memory traffic, especially in data-intensive jobs, and execution time.

**Pruning the design space:** We now study the trade-off between accuracy and execution time for approximate computing applications, and select approximate variants close to the pareto optimal curve. However, typical applications have a large number of loops which can be perforated in several ways, as well as synchronization primitives that can be elided, and data types whose precision can be lowered.

Considering all approximation possibilities makes the design space intractable, in the order of 1000s of approximation versions for typical applications [53, 12, 40, 24]. We use two ways to prune the approximation design space. First, we employ an "almost" exhaustive exploration that leverages programmer hints from the ACCEPT framework [47]. ACCEPT lists a maximum of 10 loops that can be perforated for each examined application, as well as data types whose precision can be lowered. We perforate each loop by different factors, and only preserve variants with inaccuracies lower than 5%. We follow the same process for synchronization elision, and high-precision data types. Second, for applications not supported by ACCEPT we use `gprof`, an application profiling tool to determine which functions contribute the most to execution time. In all examined applications, this corresponds to 2-4 functions, which we perforate by varying degrees. This approach also resulted in a manageable number of favorable approximate variants, consistent with those obtained using the hints from ACCEPT.

Figure 1 shows the application design space exploration for all examined applications. We use three popular cloud services, `NGINX`, `memcached`, and `MongoDB` as the latency-critical interactive applications, and 24 data mining, bioengineering, and scientific workloads as the approximate applications. For now we co-schedule each interactive service with one of the approximate applications at a time on a high-end two-socket server platform. More details on the applications and systems can be found in Section 5.

Odd rows in the figure show the tradeoff between execution time and inaccuracy across approximate variants for each approximate application. The blue dots in the scatter plots represent all examined approximate variants, the green dot represents precise execution, and the red dots represent approximate variants close to the pareto-optimal frontier, and hereafter used by Pliant. Both the number of selected approximate versions and their relative impact on performance and inaccuracy varies across applications. For example, `canneal` has four versions residing close to the pareto curve, while `raytrace` only has two. Similarly, while increasing inaccuracy has an almost inversely proportional impact to execution time, all approximate versions of `water_spatial` reside in an almost vertical line.

Even rows in the figure correspond to the impact precise execution and each of the selected approximate variants (red dots) have on the tail latency ($99^{th}$ percentile) of the three interactive services. Approximate variants are ordered from left to right in the way that they appear in the scatter plot. First, we observe that precise execution almost always leads to considerable QoS violations across interactive and approximate applications. Second, approximation has a different impact on each of the three interactive services. While switching from precise to the least approximate variant is enough for the I/O-bound MongoDB to meet its QoS in many cases, both NGINX and mem-cached exhibit higher sensitivity to resource contention. This translates to requiring the approximate application to run at its most approximate variant for QoS to be met. Even so, there are approximate applications where, despite the reduction in latency from employing approximation, the latter alone is not enough to meet QoS, e.g., *kmeans-NGINX*, *PLSA-Memcached*, and *SEMPHY-NGINX*. Conversely, there are cases, e.g., *canneal-Memcached*, and *water_nsquared-Memcached* where approximation does not have a substantial impact on tail latency. These correspond to approximate versions that do not significantly decrease contention in shared resources. In these cases reclaiming resources from the approximate application becomes necessary to meet QoS; approximation is then used as a means to avoid prolonging the application's execution time. Finally applications like `Bayesian` and `PLSA` offer a very rich design space with 8 approximate variants on the pareto curve each; this allows the runtime to sacrifice the minimum amount of quality necessary to meet QoS at each point.

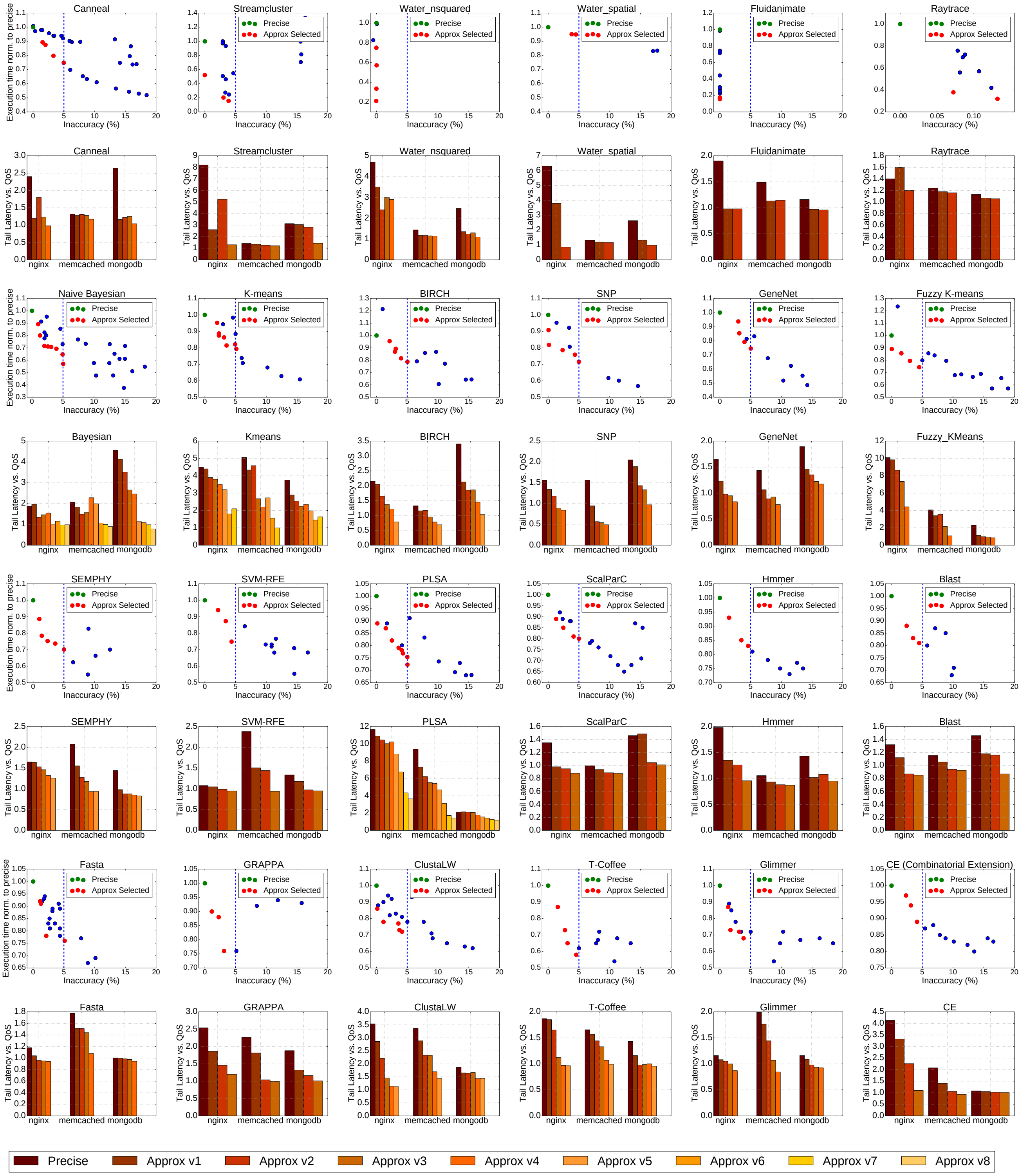

**Figure 1: Approximation design space exploration for 24 PARSEC, SPLASH-2, MineBench, and BioPerf applications. Odd rows show the trade-off between execution time and inaccuracy for different approximate variants of each application. The green dot corresponds to precise execution, blue dots correspond to all examined approximate variants, and red dots to the selected variants close to the pareto-optimal curve. The vertical line corresponds to the max permissible loss of output quality. Even rows show the impact each of the selected approximate variants has on the tail latency of the three examined interactive applications.**

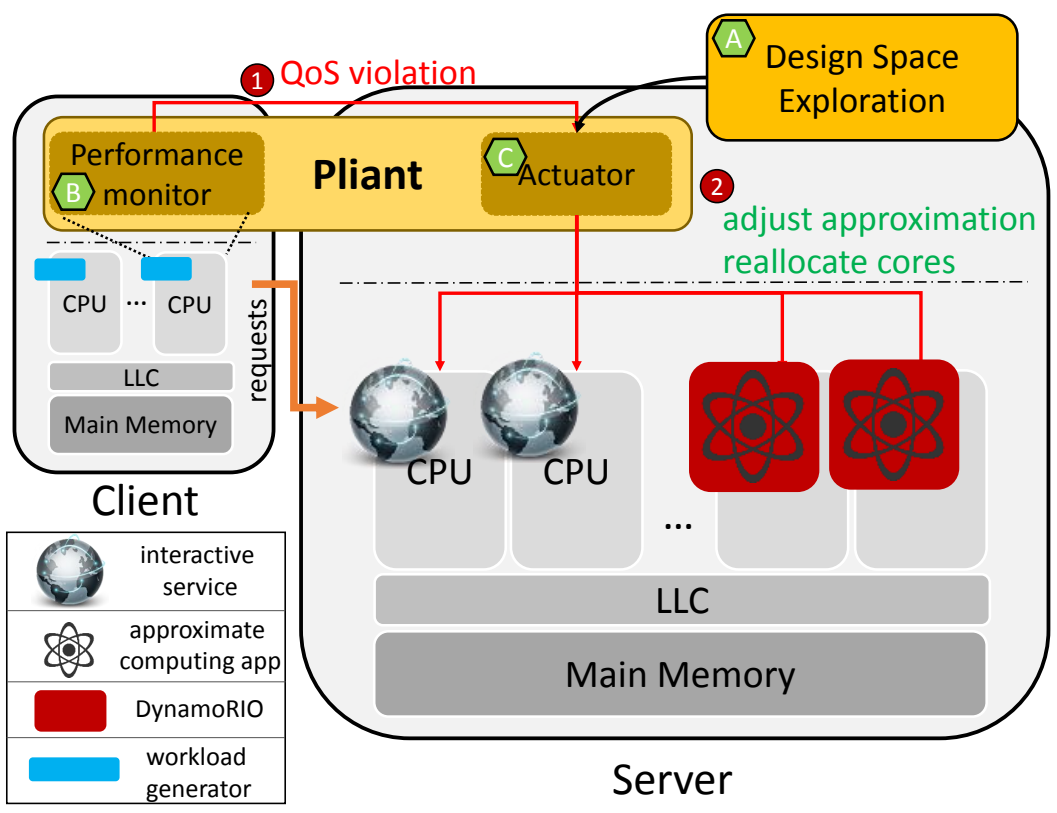

**Figure 2: Overview of Pliant's exploration and runtime components.**

# 4. The Pliant Runtime

## 4.1 Overview

Pliant consists of an *instrumentation system* that explores the approximation design space offline, and an *online runtime* that monitors performance and adjusts the degree of approximation during periods of high resource contention. Pliant's user interface involves expressing an interactive service's QoS target, and an approximate application's nominal execution time, its output quality metric, and its tolerable quality loss.

The *instrumentation system* explores the various approximation techniques described in Section 3, and obtains the ordered list of approximate variants close to the pareto frontier seen in Fig. 1. This process only needs to happen once, unless the application design changes. Having this information, the *runtime* can dynamically determine the degree of approximation needed to meet QoS at each point in time.

The runtime in Pliant consists of a *performance monitor* and an *actuator* based on dynamic recompilation, as shown in Figure 2.

The *performance monitor* is a lightweight tracing runtime that instruments the interactive applications, and continuously samples their end-to-end latency (average and tail). Since QoS metrics capture the end-to-end latency of a service, the monitor resides on the client, and is designed to not incur any measurable overhead to the interactive service, either in terms of throughput or latency. Upon detecting a QoS violation for the interactive workload, the monitor informs Pliant, which takes action via its *actuator* module. The actuator is responsible for determining the appropriate approximation variant and resource allocation, based on the monitored tail latency, and enforcing the chosen degree of approximation. Both components are designed to incur *minimal runtime overheads* from monitoring and dynamic recompilation, to be *transparent to the user*, and *preserve the performance of both the interactive and approximate application*, without exceeding the specified inaccuracy threshold.

## 4.2 Dynamic Recompilation

The Pliant actuator relies on DynamoRIO [2], a dynamic recompilation tool, to adjust an application's degree of approximation at runtime. The DynamoRIO API provides the ability to control applications at the granularity of individual instructions, as well as at the coarser granularity of functions. To avoid performance overheads from instrumentation, we use DynamoRIO at coarse granularity. Specifically, we use the `drwrap_replace()` interface to dynamically replace functions in the program with their approximate variants. Additionally, Pliant uses DynamoRIO's ability to trap Linux signals received by the application, to signal when a switch to/from an approximate function must occur.

Pliant first uses the approximate variants extracted from the design space exploration in Section 3 to construct a single application binary. This aggregates all the different versions of the functions that house the perforated loops, including one version that corresponds to precise execution. Each approximate variant is then mapped to a unique Linux signal; upon receiving the specific signal DynamoRIO switches the application to the corresponding approximate variant.

Approximate applications are executed over DynamoRIO as follows. First, DynamoRIO reads the program addresses of the precise and approximate versions for each approximated function at the start of the program. Second, during runtime, DynamoRIO traps the mapped Linux signals sent to the approximate application by the actuator. Third, when a signal is received at runtime, `drwrap_replace()` replaces the pointers to the original precise version of a function, to the corresponding approximate version using the program addresses read at start-up time. `drwrap_replace()` is also used to switch between approximate variants, or to revert back to precise execution.

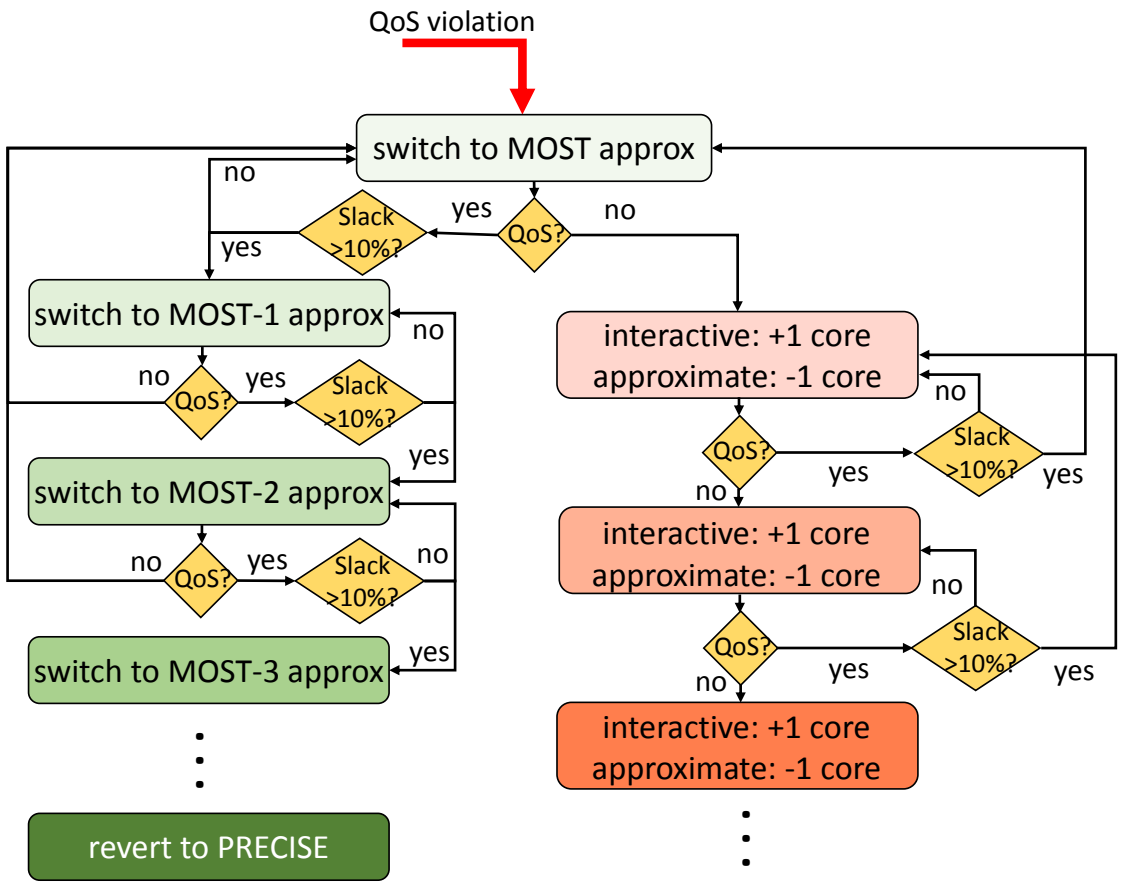

**Figure 3: Control flow of Pliant's runtime algorithm.**

Running an application over DynamoRIO can introduce overheads for the approximate applications. Across the 24 approximate applications we study, the execution time overhead is 3.8% on average, and up to 8.9% in the worst case (see Section 6). Prior work, such as ProteanCode [28] has shown that the overheads from tools like DynamoRIO are often prohibitively high for online code transformations, such as inserting non-temporal cache hints before the execution of certain loads to avoid cache contention. In that case, the high overhead of DynamoRIO comes from requiring code transformations to happen at the granularity of individual instructions. Because Pliant only switches between precise and approximate variants at coarse granularity, it can leverage dynamic recompilation with marginal overheads. Additionally, the overhead of dynamic recompilation in Pliant is almost always hidden by the shorter execution time of applications employing approximation to reduce resource contention.

### 4.3 Runtime Algorithm

Pliant uses the output of the performance monitor to determine the degree of approximation and resource allocation at runtime. Fig. 3 shows the control flow of Pliant's runtime algorithm. Initially execution starts at precise mode, and with a fair allocation of cores. In the event of a QoS violation, Pliant switches the co-scheduled application to its most approximate version to avoid prolonged degraded performance for the interactive service. If QoS is met in the next decision interval (1s by default), Pliant checks the latency slack of the interactive service. If the latency slack is greater than 10%, the runtime incrementally reverts back to less approximate versions - and potentially precise execution - to avoid unnecessarily penalizing the approximate application's output quality. If QoS is met, but there is not sufficient latency slack, Pliant remains in the same state (approximate or precise) for the next decision interval.

As shown in Fig. 1, there are cases where approximation alone is not enough to meet the interactive service's QoS. If the application runs in its most approximate variant and QoS is not met, Pliant additionally reclaims cores from the approximate application, one per interval until QoS is met. Once QoS is met, Pliant checks again for latency slack. If slack is greater than 10%, the runtime reverts to the previous state by returning the reclaimed core to the approximate application. If slack remains high, the runtime additionally decreases approximation to the minimum needed to meet QoS. Finally, if during runtime, the approximate application is operating at an approximation degree other than the highest and a QoS violation occurs, it immediately reverts to its most approximate variant.

Varying the slack threshold affects Pliant's agility in adjusting resource allocations and approximation degrees. Lowering the threshold further results in frequent ping-ponging between states, and higher overheads from DynamoRIO. Relaxing the threshold does not have an impact for the interactive service, but can degrade the performance and/or quality of the approximate application, when running with a higher approximation degree, or fewer resources than necessary. Unless otherwise specified, we use a 10% latency slack threshold.

| | |
|---|---|
| **Model** | Intel Xeon E5-2699 v4 |
| **OS** | Ubuntu 16.04 (kernel 4.14) |
| **Sockets** | 2 |
| **Cores/Socket** | 22 |
| **Threads/Core** | 2 |
| **Base/Max Turbo Frequency** | 2.2GHz / 3.6GHz |
| **L1 Inst/Data Cache** | 32 / 32 KB |
| **L2 Cache** | 256KB |
| **L3 (Last-Level) Cache** | 55 MB, 20 ways |
| **Memory** | 16GBx8, 2400MHz DDR4 |
| **Disk** | 1TB, 7200RPM HDD |
| **Network Bandwidth** | 10Gbps |

**Table 1: Platform Specification**

### 4.4 Multi-Application Colocations

So far we have assumed that an interactive service is colocated with a single approximate application. To increase utilization, servers often co-schedule multiple jobs per physical host, especially when each task is short [56]. We now extend Pliant to handle more than one approximate applications per machine. The system starts again from a fair resource allocation, and all approximate applications operating in precise mode. When a QoS violation is detected Pliant manages the approximate applications in a round-robin fashion, to avoid penalizing any of the applications in a disproportionate way. It first switches one workload (selected randomly) to its most approximate variant, and if QoS is not restored it moves to the next. If all applications operate in their most approximate variant and QoS is still not met, Pliant reclaims cores from the approximate applications, one application and one core at a time until QoS is met. The round-robin arbiter is simple, scalable, and preserves all co-scheduled applications' performance in practice (see Sec. 6); in Section 6.5 we discuss more sophisticated policies to manage multiple approximate jobs.

## 5. Experimental Methodology

**Interactive services:** We use three latency-sensitive applications, NGINX, memcached, and MongoDB.

- NGINX [5] is a high-performance HTTP webserver, and is currently responsible for 38% of all live websites as of April 2018 [6]. We use NGINX as a front-end webserver to display static HTML files. The input dataset consists of one million unique HTML files of 1KB each. The QoS target for NGINX is determined as the $99^{th}$ percentile latency before the knee of the latency-throughput curve when the application runs in isolation, and is set at 10ms, consistent with related work [17, 14, 8, 29, 22].
- Memcached [1] is an in-memory key-value store, often used as an object caching tier in cloud services [31, 29, 30]. We configure its dataset to hold 5 million items, each with 30B key, and 200B value. The QoS target for memcached is defined using the same process as above, and set to 200us, consistent with prior work [33, 17].
- MongoDB [3] is one of the most popular NoSQL databases, and is widely used in industry for back-end data storage [4, 22]. We use MongoDB 3.2.16 compiled from source, and compose a dataset with 160 million records, each with 10 fields and 100B per field. The dataset is 178GB, including indices and metadata.

All three interactive services are driven by open-loop workload generators. We instantiate enough clients to avoid client-side saturation in all cases, and ensure that the majority of latency is due to server-side delay. Unless otherwise specified, we run the interactive services at high load, approximately 75-80% of saturation.

**Approximate computing applications:** We use 24 data mining, bioengineering, and high performance computing applications from four benchmark suites as the approximate computing applications. Specifically, we use three workloads (fluidanimate, canneal, streamcluster) from PARSEC [12], three workloads (water_spatial, water_nsquared, raytrace) from SPLASH-2 [53], ten applications (Naive Bayesian, K-means, SEMPHY, Fuzzy-K-means, BIRCH, SNP, GeneNet, SVM-RFE and PLSA, ScalParC) from the Minebench benchmark suite [40] and 8 applications (Hmmer, Blast, Fasta, GRAPPA, ClustaLW, T-Coffee, Glimmer, CE) from the Bioperf benchmark suite [9]. All selected applications have metric(s) to quantify the quality of their output, and have been previously shown to tolerate some loss in their quality for improved performance and/or efficiency [49, 47, 39]. We select workloads from several fields to ensure good coverage of the approximate computing application space, and to highlight how Pliant behaves under different application characteristics.

**Systems:** We use a dual-socket, 44-physical core (88 logical core) platform, with 128GB of RAM as the server. Table **??** summarizes the specification of our experimental platform. To avoid NUMA effects, we only use one of the sockets for the interactive service, and the approximate applications. The interactive service, and the approximate workloads are instantiated in separate Docker containers, and pinned to different physical cores of the same socket. The containers share the 56MB last level cache (LLC), the main memory, and the NIC. An additional 6 physical cores are dedicated to network interrupts (soft_irq) to avoid interference with application threads. The remaining physical cores are fairly shared across the two Docker containers. For now we assume a single approximate workload co-scheduled with an interactive service. In Section 4.4 we also discuss how Pliant treats multiple approximate applications colocated with a latency-critical service. In that case the total available resources are again fairly shared among all applications at start up time.

## 6. Evaluation

We first evaluate Pliant's dynamic behavior for a few representative approximate applications, then show its performance and efficiency across all studied applications, and finally show the runtime's sensitivity to configuration parameters.

### 6.1 Dynamic Behavior

Figure 4 shows Pliant's dynamic behavior when each of the three interactive services is colocated with one of four selected approximate workloads. We select workloads that exhibit diverse characteristics with respect to their resource requirements, performance sensitivity, and number and effectiveness of approximate variants. By default Pliant uses a one second decision interval at the end of which, it makes a decision on the degree of approximation and core allocation needed for the two applications. In Section 6.4 we study the impact of varying the decision granularity. When a QoS violation is detected, the runtime switches the colocated application to its most approximate variant, and if that is not sufficient, it additionally reclaims physical cores and yields them to the interactive service, one core per decision interval. To avoid penalizing the approximate application when the interactive service has a lot of latency slack, Pliant also returns cores to the approximate workload when slack exceeds 10%.

We first examine applications in the same column. When *canneal* is co-scheduled with NGINX it almost immediately has to switch to the most approximate of its four versions, due to high compute and cache contention, and additionally relinquish 1-2 cores to the interactive service. As the tail latency of NGINX drops, *canneal* reclaims these cores, and additionally switches to an implementation variant closer to precise towards the end of its execution. Memcached experiences even higher sensitivity to resource contention, forcing *canneal* to operate in its most approximate variant for the majority of execution, and yield up to 3 cores to address short bursts of high tail latency. In contrast, the I/O-bound MongoDB needs no additional cores to meet its QoS target, and even enables *canneal* to run at precise mode for significant periods of its execution. We observe similar trends for the three interactive services, across the other approximate applications shown in Fig. 4. For example, only *raytrace* and *Bayesian* have to yield cores for brief periods of time when co-scheduled with MongoDB.

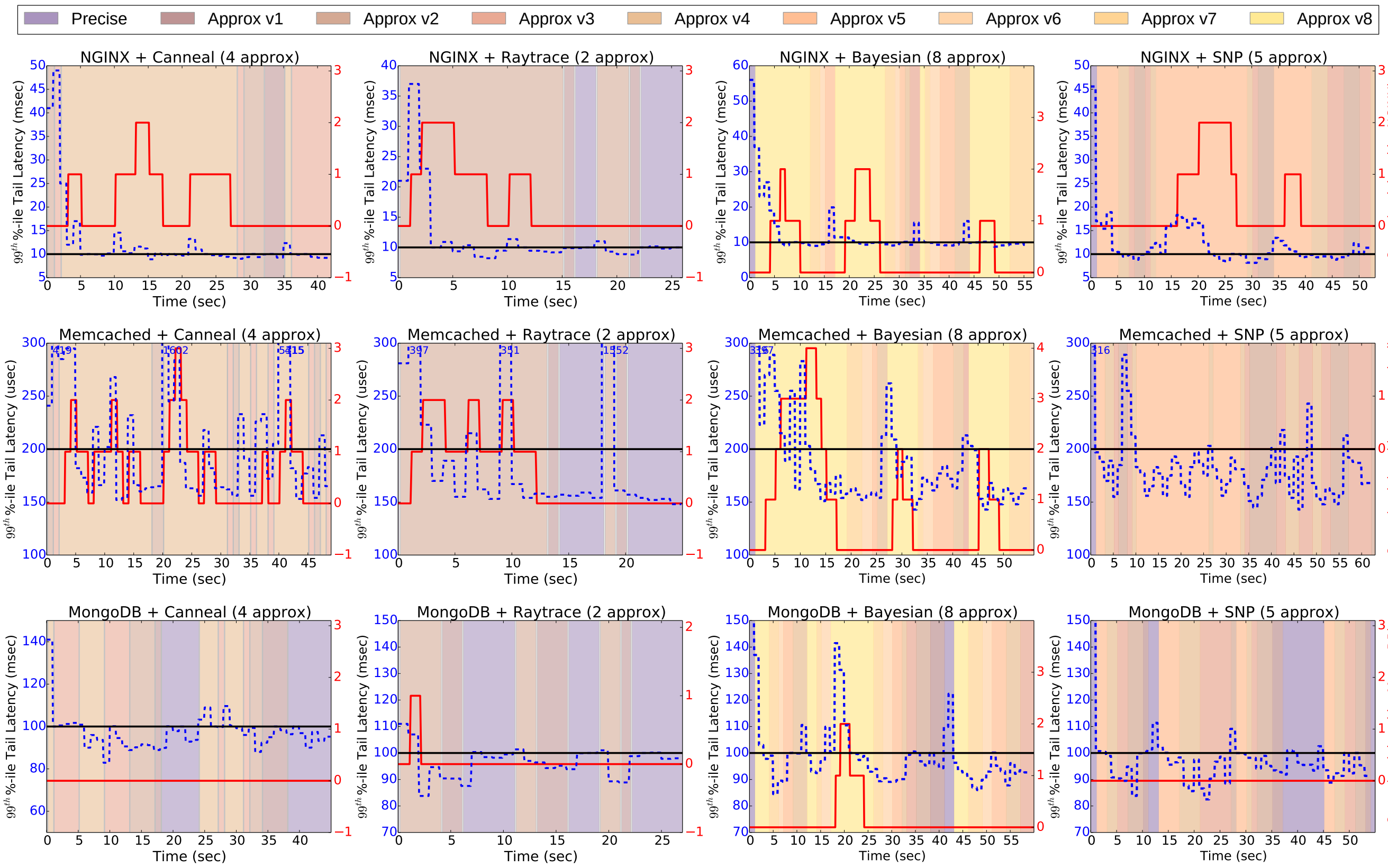

**Figure 4: Pliant's dynamic behavior when each of the three interactive services (one per row) are colocated with selected approximate computing workloads. The left y-axes show the tail latency of the interactive service over time, and the right y-axes the cores Pliant reclaims from the colocated approximate application. The black horizontal line corresponds to the QoS constraint of each interactive service. The colored background reflects the approximate version Pliant employs at each point in time. Darker colors correspond to versions closer to precise execution, and lighter colors implement more aggressive approximation, within the 5% permitted inaccuracy threshold.**

We now examine applications in the same row of the figure. Compared to *canneal*, *raytrace* only has two possible approximate variants that do not exceed the 5% allowed threshold. Because *raytrace* only introduces high compute and LLC interference in certain execution phases, it is able to leverage both its approximate variants, as well as precise execution across all three interactive services. *bayesian* offers a much richer design space with eight approximate variants close to the performance-quality pareto curve (Fig. 1). This allows Pliant to make frequent, fine-grained decisions that only sacrifice the minimum amount of output quality needed at each point in time. The figure shows that when *bayesian* is running with NGINX and it has not yielded any cores to the interactive service, tail latency is closely correlated with the approximate variant *bayesian* uses, e.g., as *bayesian* switches to decreasingly approximate versions in $t \in [26, 34]$, tail latency increases until it exceeds QoS, at which point *bayesian* returns to its more approximate version (lightest background color in the graph). Finally, *SNP* is the only of the four approximate applications pictured that enables both memcached and MongoDB to meet their QoS throughout the duration of the experiment using approximation alone. This happens because *SNP*'s approximate variants employ synchronization elision and perforation, and are particularly effective at reducing the amount of contention in the shared LLC. *SNP* only has to relinquish up to two cores when co-scheduled with NGINX.

Constraining the allocated resources does not translate to a performance penalty for the approximate applications, with all four of them achieving equal or better performance compared to precise execution, and a 2.7% average loss in accuracy.

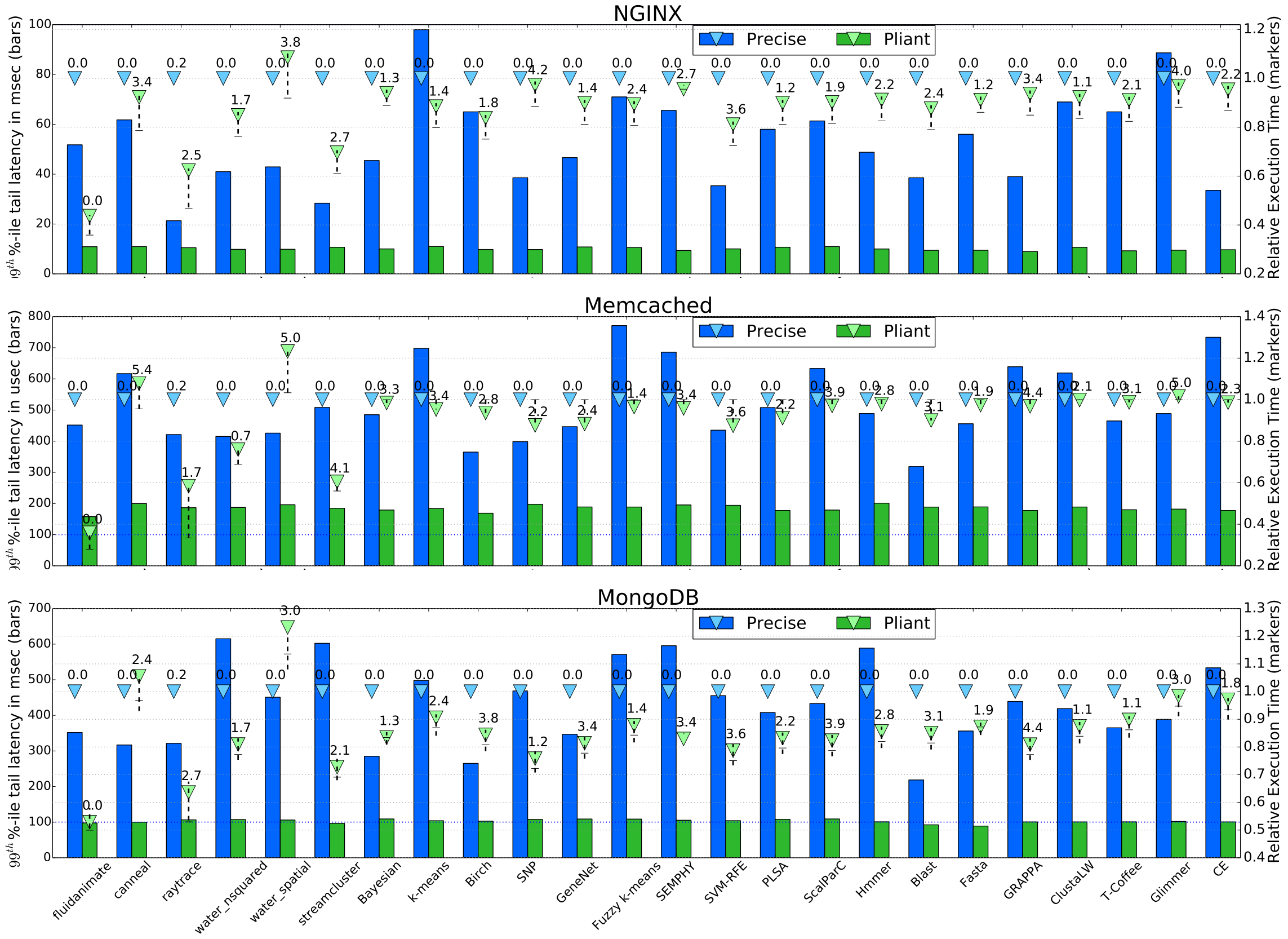

**Figure 5: Comparison of Pliant against the baseline Precise runtime for the three interactive and 24 approximate applications. The tail latency of the interactive services is shown in bars, while markers represent the execution time of approximate applications. The marker labels denote the % inaccuracy. The whiskers indicate the overhead of DynamoRIO. The QoS targets for NGINX, memcached, and MongoDB are 10ms, 200us, and 100ms respectively.**

### 6.2 Aggregate Results

We now evaluate Pliant across all 24 examined approximate applications and 3 interactive services. The decision interval is again one second. Figure 5 shows the $99^{th}-ile$ tail latency, execution time, and inaccuracy for the baseline system (Precise), and Pliant. The bars show tail latency for both runtimes, and the markers execution time for the approximate applications. The marker labels show the loss of output quality as a result of employing approximation with Pliant. The baseline precise system always operates with nominal accuracy. Additionally, in the baseline system, both the interactive service and the approximate application receive a fair resource allocation. Running in precise mode always results in severe QoS violations for the interactive service, $2.1-9.8x$ for NGINX, $1.46-3.8x$ for memcached, and $2.08-5.91x$ for MongoDB. In comparison, Pliant meets QoS for each of the interactive services across the 24 colocated approximate applications. Additionally, all approximate workloads, except for water_spatial, maintain their nominal performance (precise execution), and in several cases improve it. In the case of water_spatial, the selected approximate variants do not significantly reduce its execution time, and its decreased core allocation results in higher execution time than in precise mode. water_spatial also experiences an unusually high instrumentation overhead from DynamoRIO (seen from the whisker in Fig 5), which also contributes to its execution time.

The overhead from DynamoRIO is 3.8% in execution time on average, and up to 8.9%. The reason behind the low overhead is that dynamic instrumentation is invoked at coarse granularity, as opposed to instruction-level transformations [28]. Finally, the loss in output quality is 2.1% on average, and within the 5% tolerable limit for all applications, except for *canneal* when colocated with memcached. In that case inaccuracy is 5.4%, due to some non-determinism caused by synchronization elision.

### 6.3 Multi-Application Colocations

We now evaluate the case where Pliant handles more than one approximate computing application sharing a physical host with an interactive service. In that case Pliant examines approximate applications in a round-robin fashion to ensure

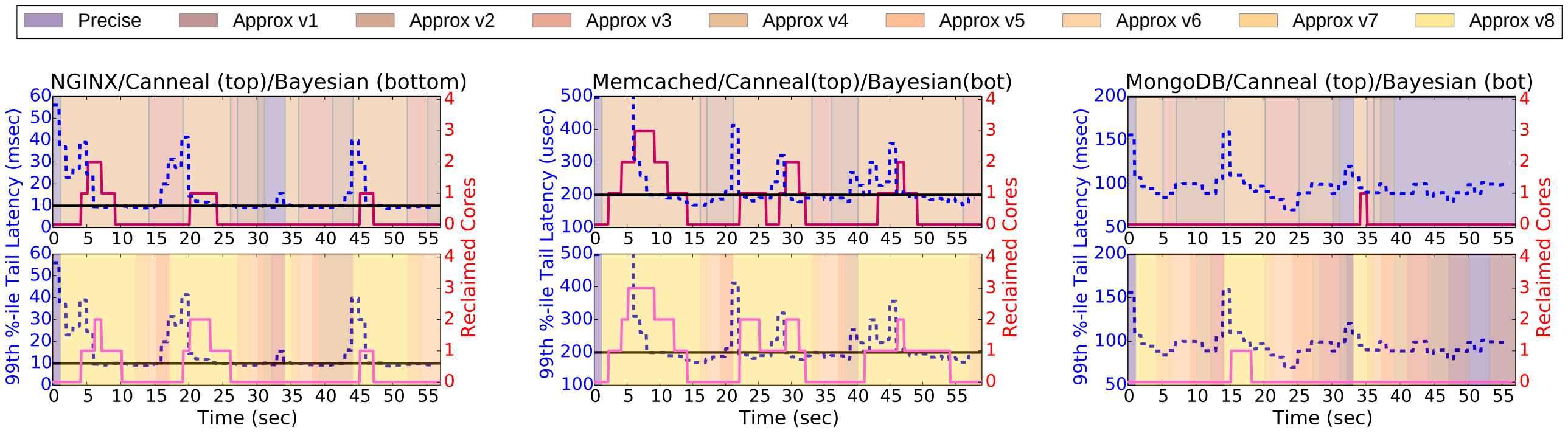

**Figure 6: Pliant managing colocations with multiple approximate applications at a time (*canneal*, and *bayesian* in this case). The top graph shows the approximate variants, and cores reclaimed from *canneal* over time, while the bottom graph shows the same for *bayesian*.**

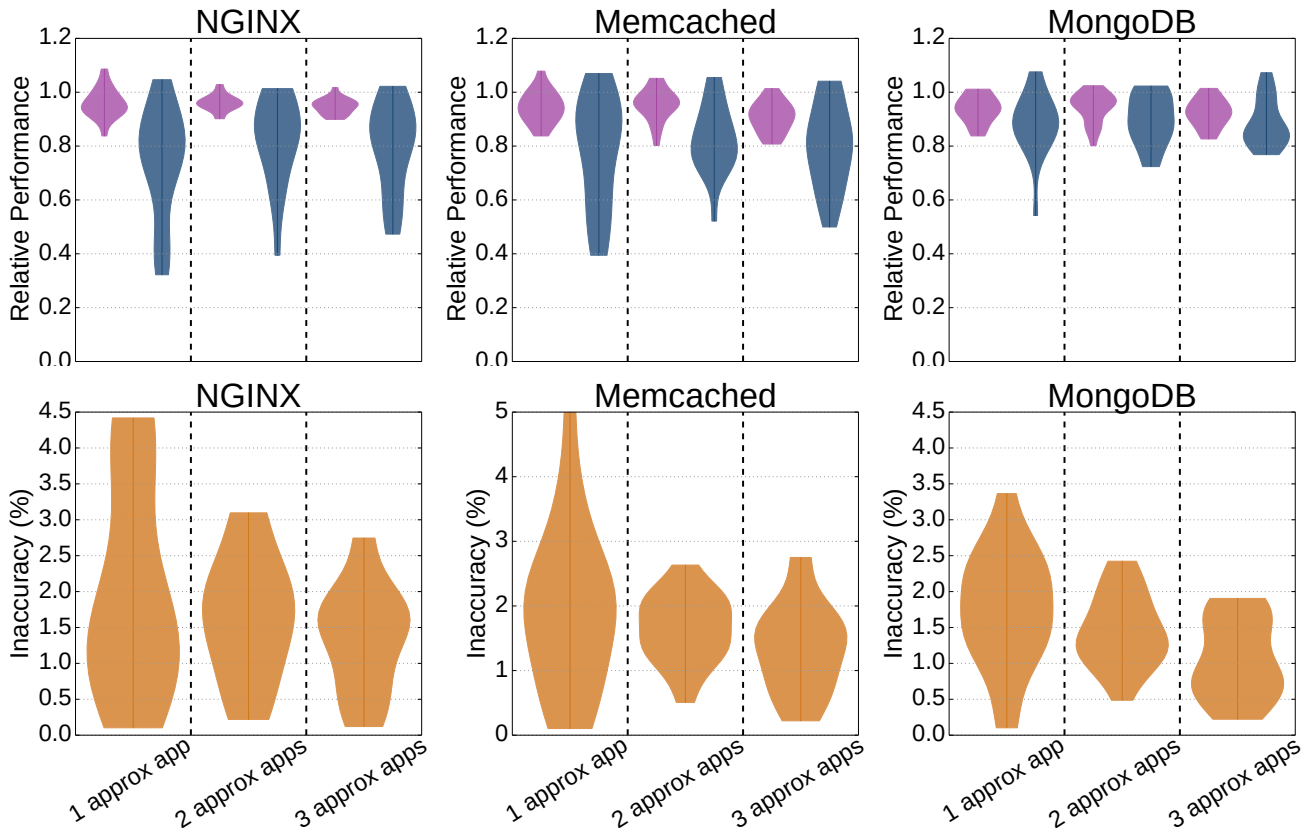

**Figure 7: Violin plots of tail latency for the interactive service (purple), execution time (blue) and inaccuracy (orange) for the approximate workload across colocations with 1, 2, and 3 approximate applications for the three interactive services. Tail latency is normalized to QoS, and execution time normalized to precise execution. The limits of the violins show the min and max value of each metric.**

that no approximate application is penalized disproportionately.

**Selected colocations:** Figure 6 shows examples of two approximate applications at a time sharing a server with each of the three interactive services. The top figure shows the approximate variants of *canneal* over time, and the cores Pliant reclaims from it to yield to the interactive service. The bottom figure shows the same metrics for *bayesian*. The interactive service's tail latency is shown in both figures. Unlike in the case where *canneal* or *bayesian* alone were colocated with NGINX, when multiple cores had to be reclaimed for NGINX to meet its QoS, now each approximate application at most yields one core to NGINX. Therefore for a large fraction of the execution, approximation alone is sufficient to meet the interactive service's QoS. This enables both approximate applications to keep their quality loss low, and preserve their nominal execution time.

As before, memcached is more sensitive to interference than the other interactive services. This results in employing more aggressive approximation variants and reclaiming a larger number of cores from both approximate workloads. In contrast, MongoDB rarely needs additional cores, while towards the end of the scenario it can meet its QoS while *canneal* is operating in precise, and *bayesian* in near-precise mode. Note that there is no case where a single application sacrifices a disproportionate amount of its accuracy and/or resources.

**Aggregate results:** We now generalize the previous experiment across all studied interactive and approximate applications. Figure 7 shows violin plots of tail latency for the interactive service (purple), and execution time (blue) and loss of output quality (orange) for the approximate workloads, when we colocate each interactive service with one, two, and three approximate applications at a time. The limits of the violins capture the min and max value of each metric. We examine all 2- and 3-way application combinations of the 24 approximate workloads. Across all three interactive services, as we increase the number of colocated applications the violins of inaccuracy become more centralized. This is consistent with Fig. 6 which shows that all colocated approximate applications sacrifice comparable amounts of their output quality. In comparison, when a single approximate application shares a node with an interactive workload it may have to sacrifice considerable quality to meet the interactive service's QoS, although without exceeding its 5% allowed threshold. The execution time violin plots for the approximate workloads reveal a similar trend of less diverse performance as consolidation increases.

Across the three interactive services, MongoDB incurs the lowest impact on the approximate workloads, both in terms of execution time and inaccuracy, since in many cases applications can continue to operate in precise mode, without an impact on MongoDB's tail latency.

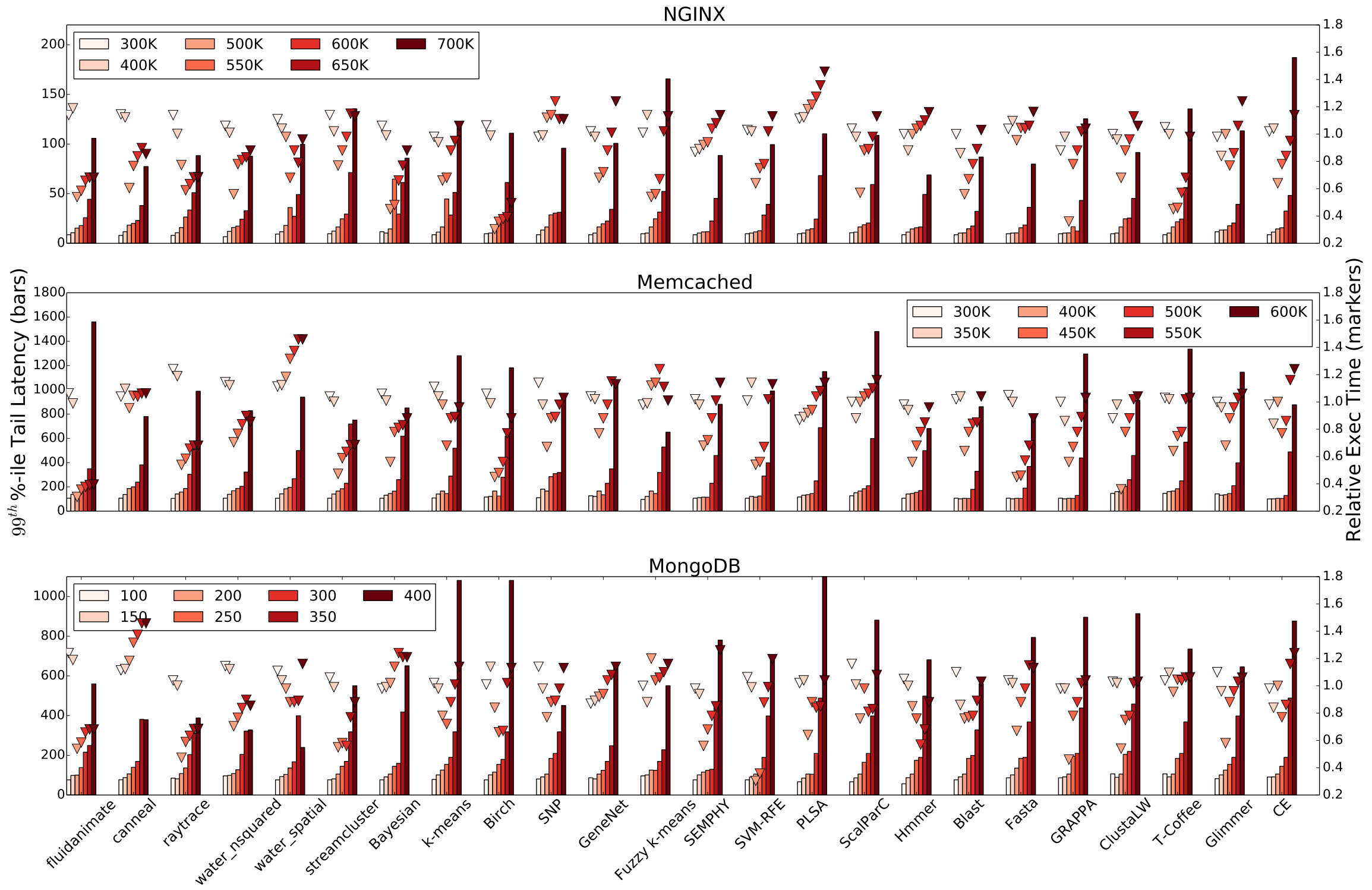

**Figure 8: Performance of Pliant across input load levels (QPS) for each of the three interactive services. The tail latency of each interactive service is shown in bars (in ms for NGINX and MongoDB, and us for memcached), while markers represent the execution time of approximate applications.**

### 6.4 Pliant Sensitivity

**Input load:** Fig. 8 shows tail latency for each interactive service, and execution time and inaccuracy for each approximate workload, as we vary the input load (QPS) of the interactive service. We focus on colocations with a single approximate workload for clarity, and examine loads between 40% to 100% of saturation in increments of 10%. Lowering the load further has no impact on either tail latency or the execution time of the colocated workloads.

When load is below 60% each of the interactive services can satisfy its QoS, while the approximate workload operates mostly in precise mode. MongoDB is an especially amenable co-runner, allowing colocated applications to operate in precise mode until it reaches 80-85% load.

For NGINX and memcached, when load is between 60-70% approximation alone is often enough to meet QoS, with memcached requiring some applications to additionally yield 1-2 cores. When load is 70-80%, approximation together with reclamation of multiple cores is needed to meet QoS, while increasing the load beyond 90% results in significant QoS violations regardless of the use of approximation. When the same applications operate in precise-only mode, QoS can only be met until 340K QPS for NGINX (48% load), 280K QPS for memcached (46% load), and 310 QPS for MongoDB (77% load). The execution time of the approximate workloads exhibits two trends. First, there are applications like water_spatial where increasing the load of NGINX results in progressively shorter execution time because the degree of approximation increases without the need for core reclamation. On the other hand, there are applications like PLSA, where increasing the load results in a slight increase in execution time because in addition to approximation, several cores must be reclaimed to meet QoS. Most applications experience both trends with execution time first decreasing, while approximation alone is used, and then increasing when cores are reclaimed.

**Decision interval:** Fig. 9 shows the tail latency and relative execution time when we vary Pliant's decision interval. The marker labels show the loss of output quality for the approximate workload. For brevity, we show a few representative approximate applications co-scheduled with `memcached`; the trend is the same for the remaining approximate workloads. When the decision interval is too coarse (above one second), the interactive service experiences prolonged QoS violations until Pliant takes action. Decision intervals of 1s or less always allow Pliant to satisfy QoS without penalizing the colocated application's execution time or accuracy beyond its allowed threshold. In theory, very short decision intervals can result in higher execution times for the approximate applications, due to frequent switching between precise and approximate versions. In practice, because Pliant is lightweight, no such degradation is observed. In fact in the case of `raytrace` there is the reverse trend; the application has higher execution time with longer decision intervals because the runtime forces it to run with a smaller number of cores than needed to satisfy QoS.

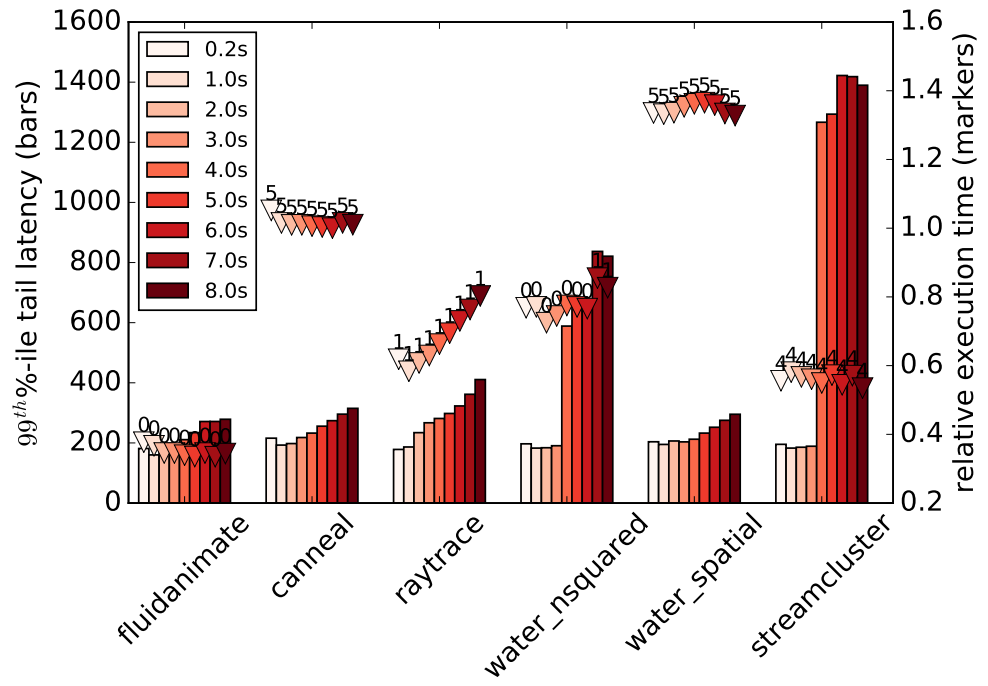

**Figure 9: Tail latency of the interactive service (`memcached`), and execution time for selected approximate workloads when varying the decision interval in Pliant. The tail latency is shown in bars, while markers represent the execution time of approximate applications. The marker labels denote the % inaccuracy.**

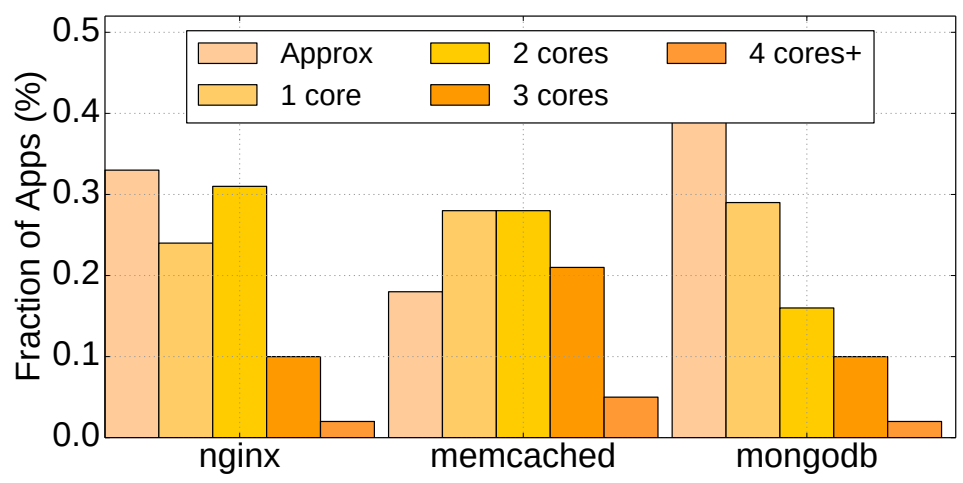

**Figure 10: Breakdown of cases where employing approximation alone is enough to meet the QoS of each interactive service, versus cases where one or more cores have to be reclaimed from the approximate workload(s).**

**Breakdown of effectiveness:** Finally, Figure 10 shows the fraction of colocations for which approximation alone was sufficient to meet the QoS of each interactive service. This includes 1-, 2-, and 3-approximate application mixes co-scheduled with an interactive service. When approximation alone is not sufficient, one or more cores are reclaimed from the approximate workloads. In the case of NGINX, for 33% of experiments approximation was sufficient to resolve any QoS violations, without constraining the resource allocation of the approximate applications. The majority of these experiments correspond to single approximate application scenarios, since in that case the interactive service starts with a larger fraction of system resources. An equally large fraction of experiments results in 2 cores being reclaimed from the approximate applications, especially for applications where approximation itself does not reduce resource contention. Reclaiming 3 or more cores is rare in practice. The results differ for memcached and MongoDB. Unlike NGINX, memcached almost always requires at least one core to be reclaimed from the approximate applications, primarily due to its strict QoS which makes it sensitive to resource interference. MongoDB on the other hand is the most amenable of the three interactive services, and can meet its QoS leveraging approximation alone, or one reclaimed core in the majority of cases. This information can be incorporated in the cluster scheduler when deciding which applications to place on the same physical node.

### 6.5 Limitations

Pliant can be extended in several ways. First, even though considering multiple approximate applications in a round-robin fashion provides a simple form of arbitration and is effective in practice, more sophisticated policies can provide better performance and/or resource efficiency. For example, considering the relative performance impact of approximation across co-scheduled applications can result in adjusting quality and/or resources from the applications that are hurt the least. We explore such policies as part of future work. Second, although we focus on core reclamation here, trading off other resources, such as cache, disk and network bandwidth, can also be beneficial, especially for network-bound analytics jobs. Finally, Pliant currently requires offline profiling to obtain the list of favorable approximate variants. This may not always be possible, especially in the context of public clouds, where the cloud provider does not have source code access to the end user's applications. In that case the user can provide the approximate variants, or hints on primitives that can be approximated using a framework like ACCEPT, and the relative impact of approximate versions can be learned at runtime [47, 38]. Alternatively, in Software-as-a-Service (SaaS), and serverless cloud settings where the user leverages the cloud's applications via fine-grained functions, the provider has source code access to perform the exploration Pliant needs.

## 7. Conclusions

We presented Pliant, a practical and lightweight cloud runtime that leverages the ability of approximate computing applications to tolerate some loss of output quality, to preserve the QoS of co-scheduled interactive services. Pliant relies on a lightweight performance monitor to track QoS violations, and a dynamic recompilation system to adjust the degree of approximation in an online and practical manner. We showed that approximation exposes a wide spectrum of operating points in terms of execution time and inaccuracy, and demonstrated that Pliant can navigate this space effectively, and preserve QoS, while using the lowest degree of approximation needed across a diverse set of interactive and approximate applications.